\documentclass[runningheads]{llncs}
\usepackage{graphicx}
\usepackage{latexsym}
\usepackage{graphicx}
\usepackage{float}
\usepackage{algorithm}
\usepackage{algorithmic}
\usepackage{enumerate}
\usepackage{multirow}
\usepackage{comment}

\usepackage{url}
\newcommand{\tabincell}[2]{\begin{tabular}{@{}#1@{}}#2\end{tabular}}

\begin{document}
\title{Automatic Description Construction for Math Expression via Topic Relation Graph}
%
\author{ Ke Yuan, Zuoyu Yan, Yibo Li, Liangcai Gao, Zhi Tang 
}
\institute{Wangxuan Institute of Computer Technology, Peking University, Beijing, 100080, China\\
 \{yuanke, yanzuoyu3, yiboli, glc, tangzhi\}@pku.edu.cn}
%
%
%
\maketitle              
\begin{abstract}
Math expressions are important parts of scientific and educational documents, but some of them may be challenging for junior scholars or students to understand. Nevertheless, constructing textual descriptions for math expressions is nontrivial. 
In this paper, we explore the feasibility to automatically construct descriptions for math expressions. But there are two challenges that need to be addressed: 1) finding relevant documents since a math equation understanding usually requires several topics, but these topics are often explained in different documents. 2) the sparsity of the collected relevant documents making it difficult to extract reasonable descriptions. Different documents mainly focus on different topics which makes model hard to extract salient information and organize them to form a description of math expressions. To address these issues, we propose a hybrid model (MathDes) which contains two important modules: Selector and Summarizer. In the Selector, a Topic Relation Graph (TRG) is proposed to obtain the relevant documents which contain the comprehensive information of math expressions. TRG is a graph built according to the citations between expressions. In the Summarizer, a summarization model under the Integer Linear Programming (ILP) framework is proposed. This module constructs the final description with the help of a timeline that is extracted from TRG.
  The experimental results demonstrate that our methods are
  promising for this task and outperform the baselines in all aspects.

\keywords{Math expression  \and Topic relation graph \and Summarization}
\end{abstract}

\section{Introduction}
Over the past decades, the web has been a rich repository of scientific information, including a large amount of technical documents, instructional documents, and academic documents.
Most of these documents contain math expressions, like formulae or equations, but understanding these math contents often remains daunting for readers~\cite{jiang2018mathematics,yuan2019automatic}, especially for junior scholars and students.

\begin{figure}
\centering
\includegraphics[scale=0.65]{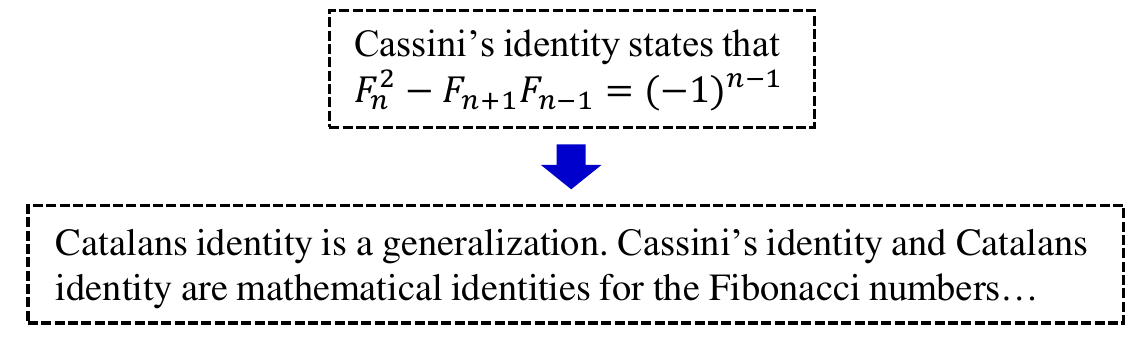}
\caption{An example. The upper box is the input which includes a math equation and its context. The below box is the constructed description for the math expression.}
\label{example}
\end{figure}

Currently, several Mathematical Information Retrieval (MIR) systems have been created to assist readers to understand the math expressions~\cite{Gao2017PreliminaryEO,wang2015wikimirs,zanibbi2016multi}. MIR systems have enabled users to retrieve information from digital collections by providing a ranked list of documents, given a math expression query. However, even the most sophisticated search engines empowered by advanced retrieval techniques lack the ability to synthesize information from multiple sources and present users with a concise yet informative response. Tools that provide timely access to, and digest of, various sources are necessary in order to  alleviate the information overload people facing. These observations have motivated us to explore feasibility to automatically construct descriptions for math expressions which can assist users to better understand the math-content in a document. Figure~\ref{example} illustrates an example. The input is the math equation ``$F_n^2-F_{n+1}F-{n-1}=(-1)^{n-1}$" and its context ``Cassini's identity states that". The output is the constructed description of the input.

However, to obtain the relevant documents which are used to construct descriptions for math expressions is challenging. Currently MIR systems perform well in identifying the documents that contain the expressions exactly matching with the mathematical query~\cite{zanibbi2016multi}. These retrieved documents mainly describe the exact topics of the query. However, they fail to retrieve the documents which contain expressions that have different structures but related information with the given query. These documents may introduce the background, supplementary or expanding information of the query. All of them could be useful for users to better understand the math expression. For instance, given a math expression     ``$P(A|B)=\frac{P(B|A)P(A)}{P{B}}$", MIR systems may retrieve the documents which mainly describe the ``Bayes Theorem". Beyond that, the documents which introduce the ``Conditional Probability" are also helpful but may not be retrieved. To address this problem, a Topic Relation Graph (TRG) is designed to collect the relevant documents that could provide the comprehensive information of the query. 

Once we obtain the relevant documents, we treat this math description construction task as a special kind of multi-document summarization task. In this type tasks, sorting the selected sentences is a challenging problem while usually ignored. However, previous linguistic study~\cite{Yan2012Timeline} reveals that the order of sentences heavily affects the understanding of text. How to reasonably order the extracted sentences to form a description is a key point in this study.

 To address these mentioned issues, we propose a  hybrid model, namely MathDes. MathDes contains two important modules: Selector and Summarizer. In the selector, a state-of-the-art MIR system is adopted to retrieve the exact topics of the math expression. Then the Topic Relation Graph (TRG) is utilized to collect the documents which describe the topics and the comprehensive information of the query. Meanwhile, the timeline information of documents is also extracted from TRG which will be used in summarizer. The summarizer selects the sentences from the documents and chronologically order them with the help of the timeline. The results show that our methods are promising for this task and outperform the baselines.

In summary, there are three main contributions: (1) we propose a novel task of automatically generating descriptions for math expressions. The constructed descriptions can help readers better understand the math-content in documents. (2) we design a Topic Relation Graph (TRG) to collect the relevant documents about math expressions, including the concepts and comprehensive information. In addition, the TRG is used to extract timeline to overcome the difficulty of organizing sentences in description construction. (3) we propose a hybrid model (MathDes) which outperforms the baselines in all aspects.

\noindent \textbf{Task Formulation}: The goal of the task is to construct descriptions for math queries.  
Formally, given a math query $q$ which contains a math expression $f$ and its context $u$, the proposed system (MathDes) will construct the textual description $M$ for the $q$. 


\section{Related Work}
\subsection{Mathematical Information Retrieval}
Mathematical information retrieval (MIR) has gained much attention in recent years, as witnessed by NTCIR conferences~\cite{aizawa2014ntcir,zanibbi2016ntcir}. MIR can be categorized into Text-based approaches~\cite{miller2003technical,sojka2011indexing}, Tree-based approaches~\cite{wang2015wikimirs,yuan2016mathematical,zanibbi2016multi,zhong2019structural} and Spectral approaches~\cite{Gao2017PreliminaryEO} according to the primitives which are used to represent math expressions. Currently, many researchers in MIR assume tree-based methods achieve better results than other methods, since tree representations can better preserve the structure and semantics of math expressions. So we adopt the tree-based MIR to obtain the exact topics of the math query in this study. 

\subsection{Document Summarization}
Document summarization has been studied intensively in recent years. Various approaches have been proposed to tackle the document summarization task, including centroid-based, graph-based algorithms~\cite{Erkan2004LexRank,Mihalcea2004TextRank}, integer linear programming (ILP) with optimization techniques~\cite{Schluter2015Unsupervised,Wan2015Multi}. Supervised model like learning to rank~\cite{yao2017content}, regression~\cite{Wan2015Multi}, and encoder-decoder neural networks~\cite{narayan2018dont,Tan2017Abstractive,yuan2019automatic} have also been leveraged to the scenario of document summarization.
Since the comprehensive information of math expressions contain timeline information, summarization paradigms utilizing timeline and temporal information~\cite{Li2013Evolutionary} are also conceptually relevant. Supervised methods have also been leveraged for timeline summarization, such as learning to rank~\cite{tran2013leveraging}, linear regression~\cite{binh2013predicting}.



\section{The Proposed Approach}
\subsection{Overview}
\begin{figure}
\centering
\includegraphics[scale=0.65]{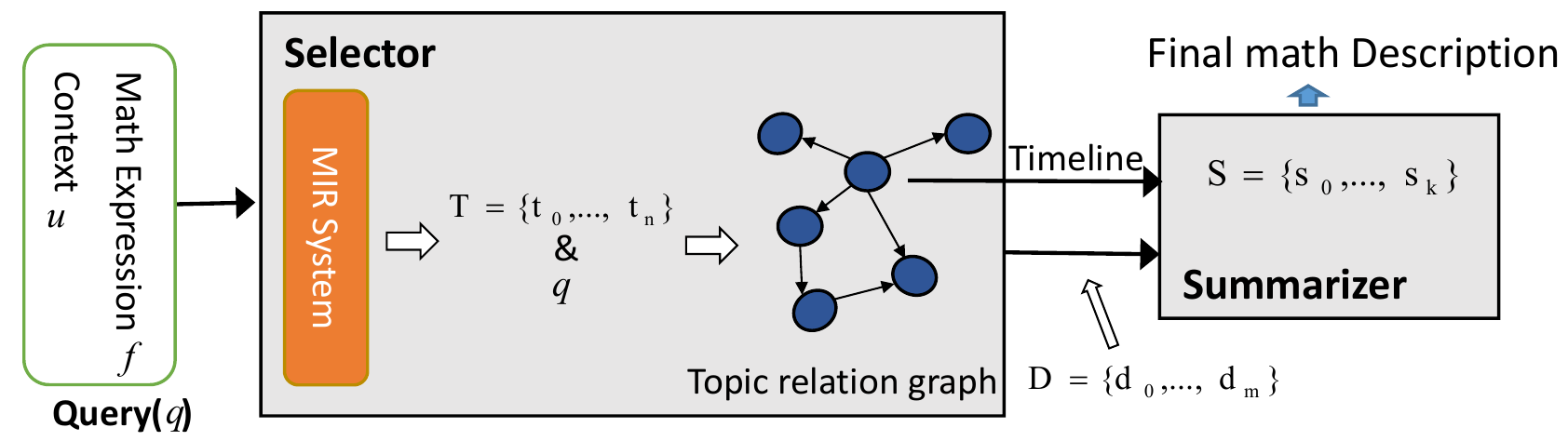}
\caption{The framework of the proposed model (MathDes).}
\label{workflow}
\end{figure}

In this section, we introduce our proposed hybrid model (MathDes).
As shown in Figure~\ref{workflow}, MathDes contains two core modules: Selector and Summarizer. The Selector is proposed to retrieve all the relevant documents $D=\{d_0,...,d_m\}$ according to the math query $q$. In the selector, a MIR system is first applied to find the exact topics $T=\{t_0,...,t_n\}$ of the query, and then these topics are treated as seeds to excavate all relevant documents($D$) with the help of the proposed \textbf{Topic Relation Graph (TRG)}. 
The Summarizer is designed to extract sentences from the relevant documents and organize them with the help of timeline that is obtained from TRG, to form the final description. 

\subsection{Selector}
\label{selector}
The Selector is designed to select all documents which are relevant to the query math expression. In the selector, there are two parts. The first one is a MIR system that is used to obtain the exact topics of the query $T=\{t_0,...,t_n\}$. The other one is processing with the help of the proposed TRG. Firstly, the retrieved exact topics are treated as seeds to seek the relevant topics which may help to explain the query math expression, then all the relevant topics are used to locate the corresponding document $D=\{d_0,...,d_m\}$ as the output of the selector. 

\subsubsection{Exact Topic Selection}
\label{sec:topic}
This part is finding the exact topics of queries. 
The input is a query and the output is the exact topics of the input. The title of a document existing in the Wikipedia\footnote{https://www.wikipedia.org/} usually denotes the topic of the document~\cite{Medelyan2008Topic}. Meanwhile, MIR systems perform well in identifying documents which mainly describe the math query. Thus, we leverage the MIR system to retrieve the documents about the query on the dump of Wikipedia and regard the titles of the retrieved documents as the topics. We adopt Wikimirs system\footnote{http://www.icst.pku.edu.cn/cpdp/wikimirs3/} to do this since it performs best in the Wikipedia math task of NTCIR-12\footnote{http://ntcir-math.nii.ac.jp/}. In this study, the top-3 retrieved topics are regarded as the exact topics of the query.

\vspace{-0.4cm}
\subsubsection{Relevant Document Selection}
\label{sec:document}

\begin{figure}[H]
\centering
\includegraphics[scale=0.5]{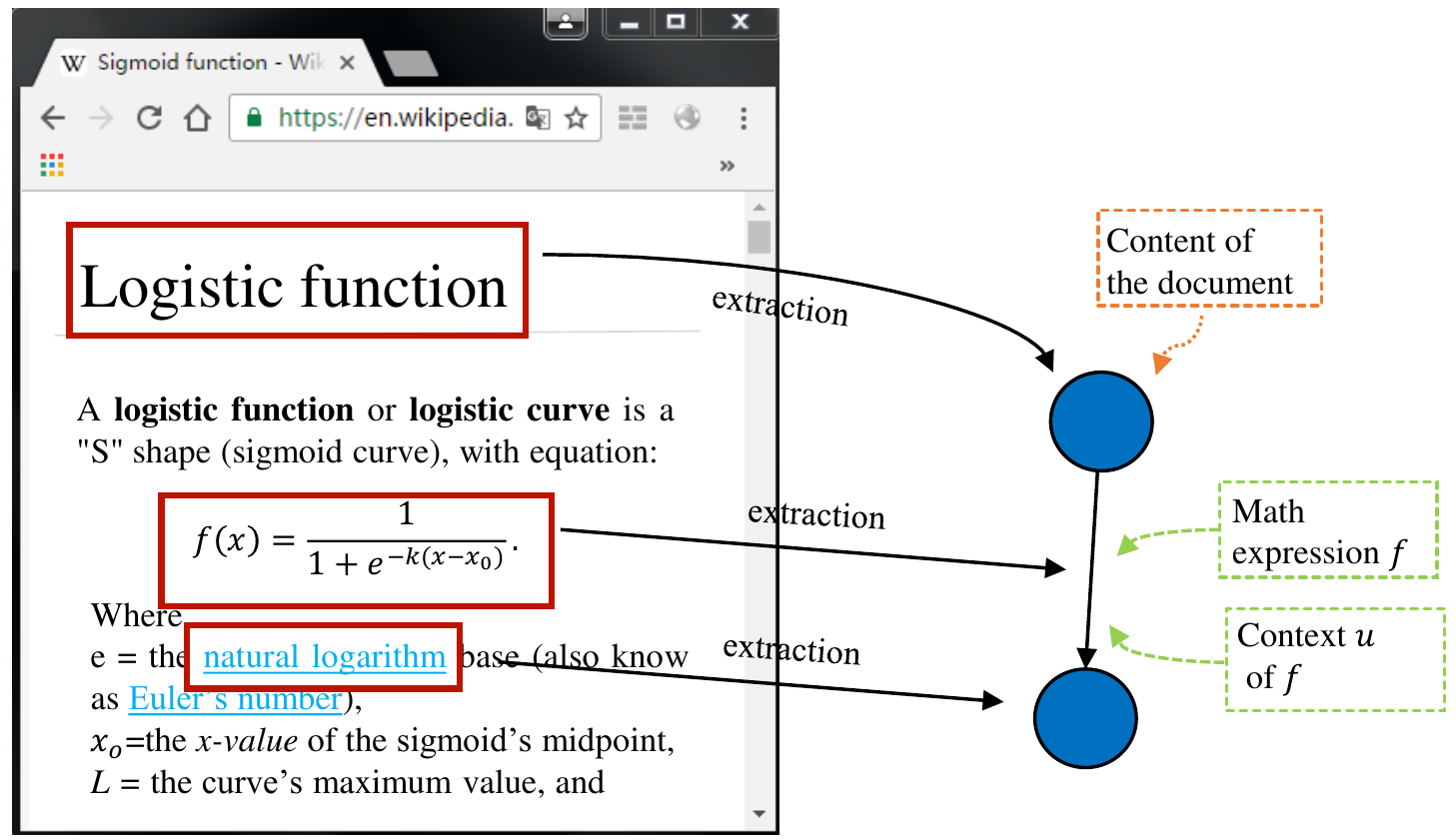}
\caption{The process of building the Topic Relation Graph (TRG).}
\label{citation}
\end{figure}

The exact topics often denote the concepts of the query, but the comprehensive information of the math expression can help users better understand it. Therefore, this part is designed to collect the documents which provide the comprehensive information of the math expression. In a document of Wikipedia, math expression usually cite other documents which provide the supportive information in its context (surrounding text). Therefore, we build a \textbf{Topic Relation Graph (TRG)}, which encapsulates the citations around math expressions. The process of building the graph is illustrated in Figure~\ref{citation}.

The vertices of TRG are the titles of documents in Wikipedia, the edges are the citations existing in the context of math expressions, and the vertices contain the title and the content of the document. The edge consists of the math expression and its context in the document.

Given the exact topics obtained from the \textit{Exact Topic Selection} of Section \ref{sec:topic} and the query, the algorithm which is used to collect the relevant documents from TRG is described in Algorithm~\ref{alg:Selectarticles}. Specifically, the most relevant documents which are connected with the seed vertices are collected as the relevant documents of the query math expression according to the Algorithm~\ref{alg:Selectarticles}. 

\begin{algorithm}
\caption{\textbf{Relevant Document Selection}.}
\small
\label{alg:Selectarticles}
\begin{algorithmic}[1] 
\REQUIRE ~~\\ 
The seed topic set, $T=\{t_1,t_2,...,t_n\}$;\\
The math expression $f$ and context $u$ of the query $q$;\\
\ENSURE ~~\\ 
The relevant document set, $D=\{d_1,d_2,...,d_m\}$;
\STATE \textbf{Initialize} $D=\emptyset$;
\label{ code:fram:extract }
\FOR{$t_i$ \textit{in} $T$}
\STATE    Find the title of document $d_j$ exactly matching with the topic $t_i$ in the \textit{TRG};
\STATE    Add the document $d_j$ into the set $D$;
\STATE    Extract outlinks $OL$ and the inlinks $IL$ document set of topic $t_i$;
\STATE $\mathop{\arg\max}_{il}\{sim(il,q)+sim(d_{il},q)\}$, $il\in IL$;
\STATE $\mathop{\arg\max}_{ol}\{sim(ol,q)+sim(d_{ol},q)\}$, $ol\in OL$;
\STATE Add the most relevant document $d_{ol}$ and $d_{il}$ which are connected by the link $ol$ and $il$ into $D$;
\ENDFOR
\RETURN $D$; 
\end{algorithmic}
\end{algorithm}

In Algorithm~\ref{alg:Selectarticles}, the similarity between the inlink/outlink edge and query is calculated as:

\begin{equation}
sim(l,q)= \cos(\vec{v}_l,\vec{v}_q)+sim(f_l,f_q)
\end{equation}
where $l$ is the inlink/outlink edge in TRG, $q$ is the query, $\vec{v}_l$ and $\vec{v}_q$ represent the averaged vectors of non-stopwords of the context existing in the edge and query, respectively. In this study, we obtain the word vectors from the \textit{GloVe}\footnote{http://nlp.stanford.edu/projects/glove/} model. $f_l$ and $f_q$ are the math expression existing in the inlink/outlink edge and the query, respectively. $sim(f_l,f_q)$ is the tree similarity between $f_l$ and $f_q$. 
We adopt the algorithm described in~\cite{zanibbi2016multi} to evaluate the tree similarity.

The similarity between the document $d$ and the query $q$ in Algorithm~\ref{alg:Selectarticles} is the cosine similarity:
\begin{equation}
sim(d_l,q)=\cos(\vec{v}_{{d_l}_b},\vec{v}_{q})
\end{equation}
where document $d_l$ is the vertex that the edge $l$ is connected with. ${d_l}_b$ is the leading paragraph of the document $d_l$, since the leading paragraph of the document in Wikipedia often gives a good brief description of the document~\cite{Peng2016News}. $\vec{v}_{{d_l}_b}$ and $\vec{v}_q$ denote the averaged vectors of non-stopwords of the leading paragraph and the context of the query $q$, respectively.

The order in which information is presented to reader critically influence the quality of a description~\cite{yao2017recent}. However, the timeline of the relevant documents is hardly extracted by the traditional methods, since timestamps of the relevant documents are not clear. Therefore, we design an algorithm to obtain the timeline by analyzing the relations existing in the TRG. The algorithm is described in Algorithm~\ref{alg:Timeline}. In this Algorithm, we assume if document $d_i$ cite document $d_j$, then document the timestamp of document $d_j$ will earlier than document $d_i$.
\begin{algorithm}
\caption{\textbf{Timeline Extraction}.}
\label{alg:Timeline}
\setlength{\belowcaptionskip}{-0.5cm}
\small
\begin{algorithmic}[1] 
\REQUIRE ~~\\ 
The relevant topics, $T=\{t_1,t_2,...,t_n\}$;\\
The related documents, $D=\{d_1,d_2,...,d_m\}$;
\ENSURE ~~\\ 
The ordering documents which are ordered by the timeline, $SD$;
\STATE \textbf{Initialize} $SD=\emptyset$;
\STATE  Remove the duplicate documents in $D$;
\label{ code:fram:extract }
\FOR{$t_i$ \textit{in} $T$}
\IF{ $t_i$ is equal to $d_j$}
\STATE $TimeStamp_{d_j}=i$;
\STATE Add tuple $(d_j,TimeStamp_{d_j})$ into the set $SD$ and Remove $d_j$ from set $D$;
\IF{$d_j\rightarrow d_k$ existing in the TRG}
\STATE $TimeStamp_{d_k}=TimeStamp_{d_j}-0.1$;
\STATE Add triple $(d_k,TimeStamp_{d_k})$ into the set $SD$ and remove $d_k$ from set $D$;
\ENDIF
\IF{$d_j\leftarrow d_k$ existing in the TRG}
\STATE $TimeStamp_{d_k}=TimeStamp_{d_j}+0.1$;
\STATE Add triple $(d_k,TimeStamp_{d_k})$ into the set $SD$ and remove $d_k$ from set $D$;
\ENDIF
\ENDIF
\ENDFOR
\STATE Sort $SD$ by the $TimeStamp$ of documents;
\RETURN $SD$; 
\end{algorithmic}
\end{algorithm}

\subsection{Summarizer}
\label{alg:Sentence}
The Summarizer is used to extract and sort sentences from the relevant documents which are collected from the Section~\ref{selector}, to construct the final math description.
We propose a summarization model under the Integer Linear Programming (ILP) framework to select the salient sentences, and then sort them with the help of timeline which is mining from TRG. The core idea of the proposed summarization model is to select sentences by maximizing the sum of weights of language concepts and the relevance with the query that appear in the final description. We formulate as:
\begin{eqnarray}
&&\max\{\sum\nolimits_{i}(w_ic_{i}+cos{(uc_{i}))}\} \\
 && s.t. \sum\nolimits_{j} l_js_j\le L \\
&& s_jO_{ji}\le c_i, \forall i,j \\
&& \sum\nolimits_j s_jO_{ji}\ge c_i, \forall i,j \\
&& c_i\in\{0,1\}, \forall i\\
&& s_j\in\{0,1\} , \forall j
\end{eqnarray}
where
$S$ is the set of sentences in the documents set.
$C$ denotes the set of bigram words\footnote{Bigrams are often used as the language concepts in ILP base summarization method.} in documents set.
$L$ indicates length limit of the description.
$u$ is the context text of the query $q$.
$w_i$ represents the weight of the bigram word $i$, the weight is frequency of the bigram word $i$.
$O_{j,i}$ denotes the indicator of whether bigram word $i$ occurs in sentence $j$.
$s_j$ is the indicator of whether sentence $i$ is selected in the description.
$c_i$ indicates the indicator of whether bigram word $j$ appears in the description.
$l_j$ represents the length of sentence $j$.
In addition, $s_j$ and $c_i$ are binary variables that indicate the presence of a bigram word and a sentence respectively. 

In our formulation, constraints (4) ensures the length of the  description not excess the length limit. Constraints (5) and (6) denote that a sentence is selected if the words in the sentence
are selected and a word is selected if at least one sentence containing the word is selected. The first part of the Equation 3 prefers to select more salient sentences while the second part prefers to select the sentences which are more relevant to the query.

After obtaining the sentences, we sort them according to the timeline which is extracted in Section~\ref{sec:topic} and the position information to form the final description. The sorting strategies are described as: (1) If the selected sentences are extracted from a single document, the selected sentences are ordered according to the order in the original document. (2) If the selected sentences exist in different documents, the sentences are sorting based on the timelines of documents. The timeline is extracted according to the Algorithm~\ref{alg:Timeline}.

Solving an integer linear program is NP-hard~\cite{Cormen:92}. However, in practice there exists several strategies for solving certain ILPs efficiently. In our study, we adapt \textit{PuLP}\footnote{http://pythonhosted.org/PuLP/}, an efficient integer programming solver, to solve our problem.


\section{Experiments}
\subsection{Data Preparation}
Two datasets are used in this paper. One is the Wikipedia, and the other one is the query dataset. Different from other domain-specific or academic corpora (e.g., ArXiv),
Wikipedia is a public available dataset and is widely used by users with different backgrounds. Thus, Wikipedia is taken as the dataset in MIR system and Topic Relation Graph (TRG). The dump size of Wikipedia is approximately 13GB compressed and 58GB uncompressed. This dataset contains 17,612,063 webpages and 676,961 mathematical expressions. The query dataset contains 82 queries consisting of the math expressions and their context, while 52 of which are described in~\cite{wang2015wikimirs} and others are used in the Wikipedia main task of NTCIR-12.
\subsection{Baseline Methods}
As a preliminary study, we only compared our system with three typical and strong traditional summarization approaches.
\textbf{Head:} The first sentence of relevant documents are extracted to construct the description. 
\textbf{Centroid:} In centroid-based summarization~\cite{Radev2000Centroid}, the centroid of a document is a pseudo-sentence which consists of words whose $tf\times idf$ scores are above a predefined threshold. 
\textbf{LexRank:} LexRank~\cite{Erkan2004LexRank} computes the sentences importance based on the concept of eigenvector centrality in a graph representation of sentences. In this model, a connectivity matrix based on intra-sentence cosine similarity is used as the adjacency matrix of the graph representation of sentences. 

In order to verfiy our proposed model, we also do some ablation studies by removing some parts of our model. \textbf{$MathDes_{w/o~timeline}$} denotes that the timeline extraction part is removed from MathDes. \textbf{$MathDes_{w/o~TRG}$} means the proposed MathDes method treats the top-n documents which are retrieved by MIR system as the source relevant documents. Please note the number of documents is the  same  in  above  methods

For fair comparison, the length of each constructed description for math expression is limited to be no more than 130 words or 5 sentences, since the average sentence length is 26 words in Wikipedia~\cite{Schluter2015Unsupervised}.
\subsection{Evaluation Metrics}
To the best of our knowledge, there are no existing gold datasets for evaluation. So we conduct manual evaluation in this study. Specifically, three volunteers who are fluent in English are participated. They are asked to sort the descriptions constructed by different methods according to two rules: 1) Whether the description can help you understand the math expression, 2) The overall quality of the description. In addition, we also evaluate the results only from the sentence extraction aspect. The volunteers are also asked to sort the descriptions on three factors: coherence, non-redundancy and overall readability, respectively. The higher ordering of the description denotes the better quality. The different types of ordering for peers are computed from the peer annotation.
\subsection{Results and Analysis}

\subsubsection{Comparison with Baseline Methods}

\begin{table}[h]
\centering
\caption{Comparison results. T-n means the constructed description rank in top-n among all constructed descriptions. The value is the average ratio of the constructed descriptions rank top-n in all the descriptions which are constructed by one method.
  }
\begin{tabular}{|c|c|c|c|}
\hline
  \textbf{Method} & \textbf{T-1}(\%) & \textbf{T-2}(\%) & \textbf{T-3}(\%) \\
  \hline
  \hline
  Head& 13.8 & 21.1&29.2\\
  \hline
  Centroid & 11.4 & 21.1&38.2\\
  \hline
  LexRank & 21.1 & 37.4&56.1\\
  \hline
  \hline
  $MathDes_{w/o~TRG}$&15.4&27.6&45.5\\
  \hline
  $MathDes_{w/o~timeline}$&30.1&59.4&77.2\\
  \hline
  MathDes&\textbf{39.8}&\textbf{63.4}&\textbf{79.7}\\
  \hline
\end{tabular}
  \label{tab:baseline}
\end{table}

Table~\ref{tab:baseline} shows the comparison results of different methods. The proposed model (MathDes) achieves the best results among all metrics. MathDes leverages Topic Relation Graph (TRG) to collect all relevant documents, and then applies the timeline to sort the selected sentences to form the description. Compared to MathDes, the $MathDes_{w/o~timeline}$ method only leverages the TRG to collect relevant documents without using the timeline. The $MathDes_{w/o~TRG}$ method only utilizes the proposed summarization method and treats the top-n documents which are retrieved by MIR system as the source relevant documents.
The Comparison between $MathDes_{w/o~TRG}$ and $MathDes_{w/o~timeline}$ shows that the TRG can collect more relevant information which can help users better understand the query math expression. Another comparison between MathDes and $MathDes_{w/o~timeline}$ demonstrates that the timeline of documents extracted from the TRG is useful to organize the selected sentences.

The leading sentences of documents are informative but merely used, since they are usually the description of the document, not the specific description for the query math expression. Therefore, Head method does not achieve competitive results. Neither LexRank nor Centroid methods perform well in our settings. These methods are believed suitable for single document summarization tasks which can not deal with the sparsity of the collected documents effectively.

\subsubsection{Readability Assessment}
The discourse coherence in the multi-document summarization task is a challenging problem. Unlike the traditional multi-document summarization tasks, where multiple documents describe one topic, the collected documents used to construct the description for math expression describe different topics. Therefore, discourse coherence becomes more serious in this scenario. Table~\ref{tab:read} illustrates the results of the manual evaluation on coherence, non-redundancy and overall readability.

\begin{table}[h]
\centering
\caption{Manual readability ordering. The values are the average ratio of the constructed description rank top-1 in all the descriptions which are constructed by one method.}
\begin{tabular}{|c|c|c|c|}
\hline
  \textbf{Method} & \textbf{Coherence} (\%) & \textbf{Non-Redundancy} (\%)& \textbf{Readability} (\%) \\
  \hline
  \hline
  Head& 22.0& 26.0&34.1\\
  \hline
  Centroid & 14.6 & 31.7&13.8\\
  \hline
  LexRank & 28.5 & 48.0&17.9\\
  \hline
  \hline
  $MathDes_{w/o~TRG}$&15.4&51.2&21.1\\
  \hline
  $MathDes_{w/o~timeline}$&30.1&60.2&35.2\\
  \hline
  MathDes&\textbf{40.7}&\textbf{64.2}&\textbf{42.3}\\
  \hline
\end{tabular}
  \label{tab:read}
\end{table}

From the results, we can find that the methods under the ILP framework perform better than other summarization methods in non-redundancy, which demonstrates ILP frameworks which are designed in this study is useful to avoid the redundancy. The difference between readability and the overall quality shows that the readability will effect the overall quality for a certain level. Meanwhile, we can observe that our methods outperform other methods among all the evaluation metrics.
\subsubsection{Error Analysis}
In this preliminary study, the proposed model (MathDes) is used to generate descriptions for queries. 
The documents collected from the TRG are relevant to the query or not, heavily affects the quality of the constructed description. In this study, the relevant documents are collected according to the topics that are retrieved by the MIR system. Therefore, the quality of the topics directly decides the quality of the final descriptions. However, MIR system may retrieve some non-relevant topics about the query, which leads to a bad quality descriptions.
This issue is highly non-trivial and has not been well addressed in the method we explored in this paper.
We leave it for further study.
\subsubsection{Case Study}
Table~\ref{tab:casestudy} shows some descriptions which are constructed by different methods. From the results, we can find that the description which is generated by Head method contains the same sentences, since there are existing duplicated documents collected by TRG. In the relevant document selection, two different topics may cite one document in TRG. Although the cluster of the relevant documents contains the same documents, the proposed method can also construct a better quality description for the query. In addition, the method incorporates the timeline information performs better than other methods.

%
%
%
\begin{table*}[h]
\centering
\caption{Descriptions constructed by different methods.}
\begin{tabular}{c|p{3cm}|p{5cm}}
\hline
  \textbf{Query} & \textbf{Method} & \textbf{Description} \\
  \hline
  \multirow{6}*{\tabincell{c}{Math expression:\\$a^2+b^2=c^2$\\ Context:\\Pythagorean theorem}}
  &Head&\footnotesize In mathematics, the Pythagorean theorem, also known as Pythagorass theorem, is a fundamental relation in Euclidean geometry among the three sides of a right triangle....\\
  \cline{2-3}
  &Centroid&\small The posterior is then generalized Dirichlet with parameters being the number of balls, and the number of marbles, in each box.... \\
  \cline{2-3}
  &LexRank&\small In equations, (this is sometimes known as the right triangle altitude theorem) where a, b, c, d, e, f are as shown in the diagram....\\
  \cline{2-3}
 & $MathDes_{w/o~TRG}$&\small The two different methods for determining the area of the large square give the relation between the sides....\\
  \cline{2-3}
  &$MathDes_{w/o~timeline}$&\small Ptolemys theorem. In mathematics, the Pythagorean theorem, also known as Pythagorass theorem, is a fundamental relation in Euclidean geometry among the three sides of a right triangle....\\
 \cline{2-3}
  &MathDes&\small In mathematics, the Pythagorean theorem, also known as Pythagorass theorem, is a fundamental relation in Euclidean geometry among the three sides of a right triangle....\\
  \hline
    \end{tabular}

  \label{tab:casestudy}
\end{table*}

\section{Conclusion and Future Work}
In this paper, we explore a preliminary challenging task that is automatically construct descriptions for math expressions. We also proposed a hybrid model (MathDes) to address this task. MathDes can tackle the sparsity issue. Meanwhile, a timeline extraction algorithm is designed based on the proposed Topic Relation Graph (TRG) to make the constructed description more fluently. The results show that our proposed method is appropriate for this task and outperforms the baselines. In addition, our method can be trivially adapted to the chemical diagrams, tables as well.

As a preliminary work, our method is under sentence extractive framework. We would like to extend our system to produce descriptions for math expressions beyond pure sentence extraction.

%
%
%
%
%
%
\bibliographystyle{splncs04}
\bibliography{acl2018}
\end{document}